\documentclass[10pt,journal,compsoc]{ IEEEtran}
\usepackage{comment}
\usepackage{color}

\ifCLASSOPTIONcompsoc
  \usepackage[nocompress]{cite}
\else
  \usepackage{cite}
\fi
\usepackage[
	colorlinks=true,
	urlcolor=blue,
	linkcolor=green
]{hyperref}

\usepackage{float}

\ifCLASSINFOpdf
   \usepackage{graphicx}
   \DeclareGraphicsExtensions{.pdf,.jpeg,.png}
\else
\fi
\begin{document}
%
\title{Crushing the Wave - new Z-Wave vulnerabilities exposed}
%
%
%
%

\author{
\IEEEauthorblockN{Noureddine Boucif\IEEEauthorrefmark{3},  \thanks{\IEEEauthorrefmark{3}boucif.nourreddine@gmail.com}}
\and
\IEEEauthorblockN{Frederik Golchert\IEEEauthorrefmark{1}, \thanks{\IEEEauthorrefmark{1}it.golchert@web.de}}
\and
\IEEEauthorblockN{Alexander Siemer\IEEEauthorrefmark{2}, \thanks{\IEEEauthorrefmark{2}alexander.siemer@gmail.com}}
\and
\IEEEauthorblockN{Patrick Felke\IEEEauthorrefmark{4}, \thanks{\IEEEauthorrefmark{4}patrick.felke@hs-emden-leer.de}}
\and
\IEEEauthorblockN{Frederik Gosewehr\IEEEauthorrefmark{5} \thanks{\IEEEauthorrefmark{5}frederik.gosewehr@hs-emden-leer.de} \\
\vspace{1em}
\IEEEauthorblockA{\textit{Department of Computer Sciences and Electrical Engineering} \\
\textit{University of Applied Sciences Emden / Leer}\\
Emden, Germany}
}}

\IEEEtitleabstractindextext{%
\begin{abstract}
This paper describes two denial of service attacks against the Z-Wave protocol and their effects on smart home gateways. Both utilize modified unencrypted packets, which are used in the inclusion phase and during normal operation. These are the commands \emph{Nonce Get}/\emph{S2 Nonce Get} and \emph{Find Nodes In Range}. This paper shows how both can be manipulated and used to block a Z-Wave gateway's communication processing which in turn disables the whole Z-Wave network connected to it.  

\end{abstract}

}

\maketitle

\IEEEdisplaynontitleabstractindextext

%
\IEEEpeerreviewmaketitle

\IEEEraisesectionheading{\section{Introduction}} 

%
%
%
%

\IEEEPARstart{W}{ith} the current trend towards industrial and private digitization, the building automation- and smart home sector have become fast growing industries, since the demand for such digitization has seen a steady increase in the last years \cite{StatistaSHR2019}. As these smart home systems are becoming more commonly used, the internal and external security of such systems is getting  more and more crucial. For this reason this paper shows an analysis of the Z-Wave smart home protocol and its implementation in regards to its security. Z-Wave is a proprietary standard, which today is owned by \emph{Silicon Labs} with the former owners being \emph{Zen-Sys} and \emph{Sigma Designs}. While most of the protocol standard, especially the security aspects, are being kept secret, previous works e.g. \cite{BehrangGhanounBH2013} have shown, that the "security through obscurity" approach is ultimately doomed to fail. While Silicon Labs customers need to implement both the provided hardware, a Z-Wave SoC (System on Chip, \cite{ZwaveSoC}) or module \cite{ZWaveModule} and the proprietary software protocol stack library and are also legally bound to a nondisclosure and confidentiality agreement regarding the secret details of the protocol, to keep the standard secure, programmers as well as attackers have shown that the protocol can be reversed engineered anyway, which lead to projects like the Open Z-Wave library \cite{OpenZWave}. With more than 2600 different Z-Wave certified products currently available, which are being manufactured by approx. 700 different companies and a U.S. market share in the home security area of approx. 90\% \cite{ZWaveAllianceReport}, the protocol is besides Zigbee the most commonly used one for home automation, which makes it interesting for but not only security researchers. 

The following analysis focus laid on finding security vulnerabilities, which could be exploited using Software Defined Radio (SDR) to send fake messages. A \emph{HackRF One}\cite{HackRF} acted as transmitter and receiver, which was controlled by GNU Radio\cite{GNU}. The Python module \emph{Scapy-radio}\cite{Scapy-Radio} was included into attack scripts for decoding and encoding packages with the goal of finding unencrypted messages. These messages are part of the protocol's design and therefore intended, but have the potential to be misused to cause unwanted system states or to control devices.
The result of this research are two novel Denial of Service attacks (DoS), which overload the Z-Wave gateway with relatively little effort. A gateway in this state will no longer process events from connected devices or the smartphone app, which disables the entire smart home network for all participants.

\section{State of the Art}

Previous researchers like e.g. \emph{PenTestPartners} \cite{Pentestpartners} discovered a  downgrade attack against the newer version of the \textit{Z-Wave} security standard. They did this within a test case using a locked door in inclusion mode while manipulating the \textit{NodeInfo} package and exploiting its backwards compatibility.
Before this attack, a security evaluation from Behrang Fouladi and Sahand Ghanoun \cite{SecurityEvaluation} focused on controlling a door lock without exploiting the default key of \textit{Z-Wave} devices, by exploiting a missing validation in the key exchange protocol handler. They were able to reset the shared network key, which was ultimately a implementation error of the door lock manufacturer. Since the implementer is capable to fit a Z-Wave system to his needs, the protocol security measures can be circumvented if implemented the wrong way. Another class of attacks target the Z-Wave routing protocol, e.g. so called Black Hole attacks\cite{BADENHOP2017112}, utilizing bad design in the routing protocol. While most of the other publications focus on decrypting messages and/or controlling the Z-wave components itself, we focused our evaluation on possible denial of service attacks like the ones shown in \cite{BADENHOP2017112}. The novel approach of our attack thereby lays in targeting the Z-Wave gateway itself and therefore the main communication hub, not the routing devices of the Z-Wave network. 

\mbox{\\}

The paper is structured as follows: in Section 3 the packages used for the attacks are explained with their intended use. Section 4 explains the methods and the procedure of the attacks including the structure of the used packages. In Section 5 the results of the attacks are depicted which are then discussed in Section 6. 

\section{The Z-Wave Protocol }
The information about the Z-Wave protocol were gathered through the various specifications of Silicon Labs which are available in the internet and through testing. Z-Wave is still a closed protocol which added to the difficulty of the research.\cite{spec}
\subsection{Nonce-Get}
\begin{figure}[ht]
\includegraphics[width=\linewidth]{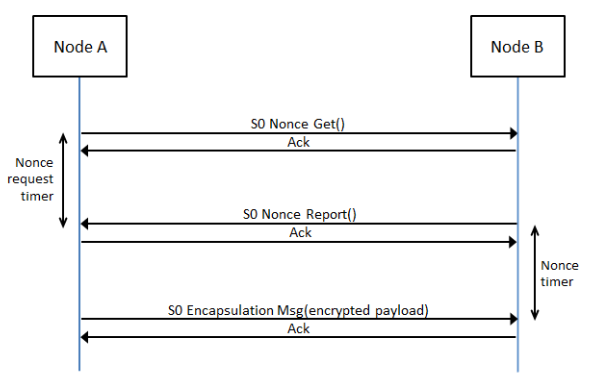}
\caption{Z-Wave-Transport-Encapsulation \cite{Z-Wave-Transport-Encapsulation}}
\label{fig:Nonce}
\end{figure}

The \emph{Nonce-Get} command is used to request a unique random number, which must never be used again. Such number is called \emph{Nonce} which is a abbreviation for \emph{Number used once}. These nonces get requested as part of the  encryption algorithm \textit{S0 Encapsulated} Messages (see figure \ref{fig:Nonce}). The initialization vector (IV) used for such encryption is spilt into two parts, namely the Nonce of A and B, which are concatenated to create the IV (IV = (nonce sender $||$ nonce receiver))\cite{ZWaveApplication}. The Sender, here \emph{Node A}, initially transmits a \textit{Nonce Get} command, which will be answered by the receiver, Node B, with an \textit{Nonce Report} package, containing the newly created Nonce from \emph{Node B}. \emph{Node A} is now able to combine both nonces as IV for the encryption of the \textit{S0 Encapsulated Message} payload. 
\newline This algorithm stays basically the same for the \textit{S2 Encapsulated Message} except for the \emph{Nonce Get} command which is \textit{S2 Nonce Get}. A connected node has to answer this request whenever another node or the gateway sends this command, but it must not, if the command is sent by multicast. The node itself isn't allowed to send it via multicast either. It should also be mentioned that, if no acknowledgement (ACK) follows as reaction, an attempt is made to route it to the receiver. But there is no rule in place stopping nodes from attempts to send a) Nonce Report packets to addresses which haven't been assigned to a node or b) to their own node address.

\subsection{Finding Z-Wave nodes in range}
\begin{figure}[ht]
\includegraphics[width=\linewidth]{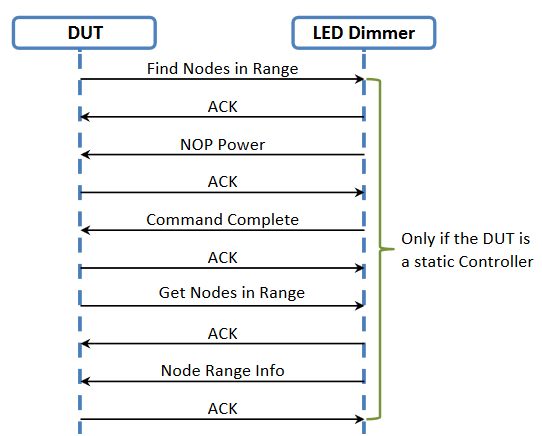}
\caption{Procedure of Find Nodes in Range command\cite{findNodes}}
\label{fig:finde}
\end{figure}

The \textit{Find Nodes In Range} command is used by the Gateway in the inclusion phase of a new devices. The device, which is being included into the Z-Wave network, will send \textit{NOP Power} packages to every address given in the command. Afterwards the device waits a moment after each packet sent to get an ACK message. Using this method the requesting device is able to discover all other devices within its range (see figure \ref{fig:finde}). If the command is completed the device sends a \textit{Command Complete} package to notify that it has finished. The command \textit{Find Nodes In Range} will generally be send  by the gateway to an device which is getting included. Devices, which aren't in the inclusion phase, should not accept this command.

\section{Methodology}
In general Python scripts have been used to create and process packets with Scapy-Radio. The Scapy-Radio version of BastilleResearch\cite{Scapy-Radio} was used for our purpose. These Packages are then processed through software defined radio with GNU Radio. Therefore the open-source Software Defined Radio HackRF One\cite{HackRF} was used as an transmitter and receiver. Two of them were needed, because they aren't full-duplex.  We had to change the GNU Radio flow graph to get at least the receiving path of 100 Kbit/s and 40 Kbit/s running. With the used flow graph we were able to send packages with 40 Kbit/s.

\subsection{Used tools: Scapy-Radio}
\textit{Scapy-Radio}, a modified version of the python program \textit{Scapy}\cite{Scapy}, which has been altered to process wireless protocols, is used to send, sniff and filter the \textit{Z-Wave} packages. The specific version used has been modified by BastilleResearch\cite{Scapy-Radio} as testing tool for \textit{IoT} radio networks and includes some Z-Wave protocol capabilities.

\subsection{Used tools: Gnu-Radio}
\begin{figure*}[ht]
\includegraphics[width=\linewidth]{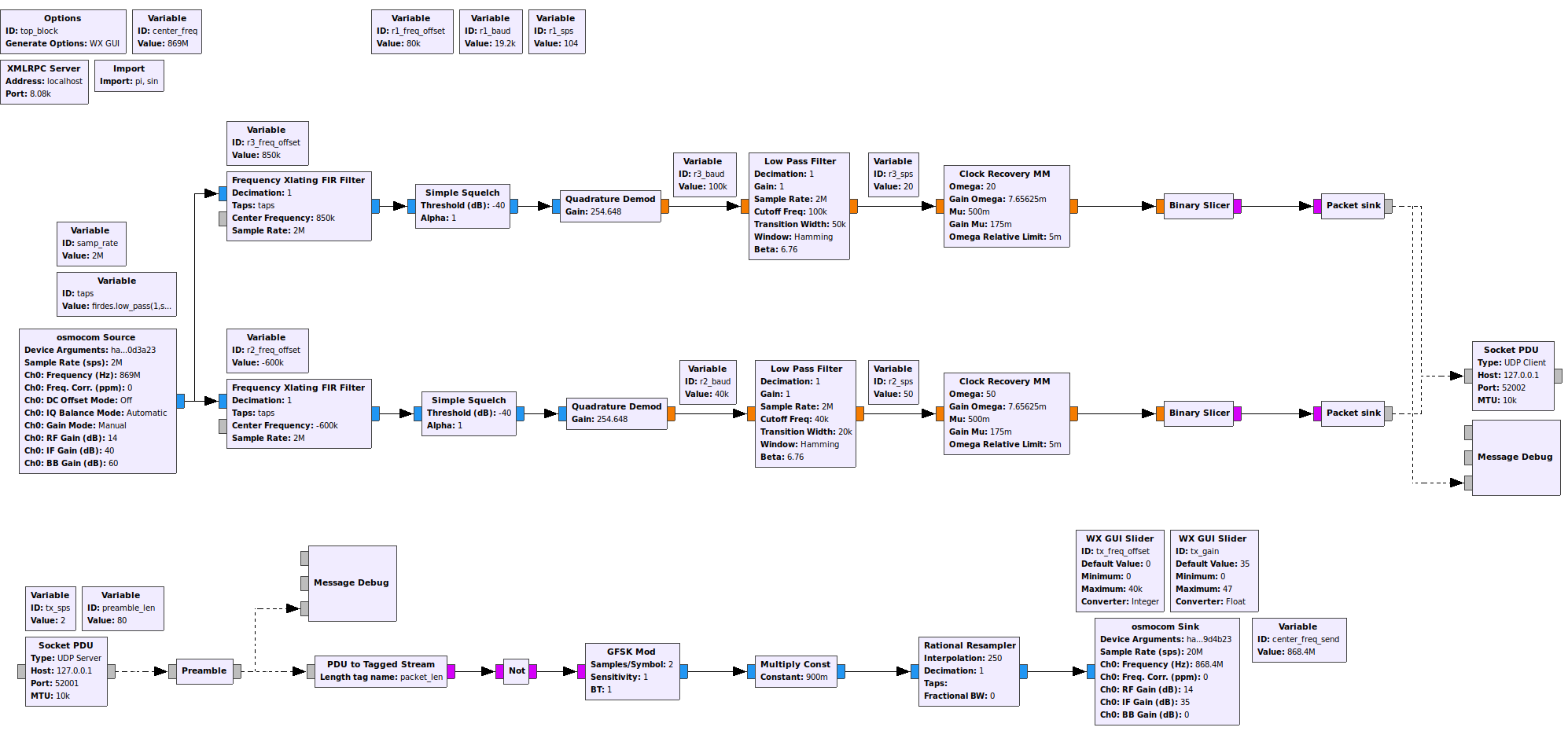}
\caption{Z-Wave Flow Graph}
\label{fig:zwave-flowgraph}
\end{figure*}

GNU Radio\cite{GNU} is a free tool for implementing software defined radios (SDR). GNU Radio uses two HackRF One in this project, one as receiver and one as transmitter. GNU Radio has a graphical user interface which can be used to create a or multiple flow graphs from blocks via drag and drop (much like Matlab Simulink). For the HackRF One a Osmocom\cite{Osmocom} source block is used. The created flow graph defines the decoding, demodulation and the general processing of the signals. The used flowgraph is shown in \ref{fig:zwave-flowgraph}.
At the end of the flowgraph the data stream is send via the system's internal loopback interface to the written Python script for further processing.

\subsection{Used tools: HackRF One}
The HackRF One\cite{HackRF} is an open-source SDR solution. This device can transmit and receive radio signals from one megahertz to six gigahertz. It can be connected via USB and be used for instance in GNU Radio or be programmed as a stand-alone solution.


\subsection{Packet Manipulation}
In search for vulnerabilities within the Z-Wave protocol or specific implementations malicious data packages were generated using Scapy-Radio. Primarily, unencrypted commands were tested as these can be used independently from the encryption and security level, which were manipulated in different ways. First the addresses were changed and combinations were tested which include single- and multicast traffic. After these tests the payload was also altered, especially the packets with the requirement of the receiver to send a answer. During these tests the commands \emph{Nonce Get / S2 Nonce Get} and \emph{Find Nodes In Range} were showing unexpected effects. Through these effects both commands could be used for a Denial of Service attack against the Z-Wave gateway.

\subsection{Nonce-Get manipulation}
\begin{figure}[h]
\includegraphics[width=\linewidth]{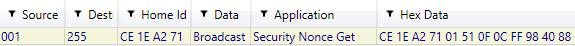}
\caption{Structure of manipulated S0 Nonce Get packet}
\label{fig:Nonce_S0}
\end{figure}

\begin{figure}[h]
\includegraphics[width=\linewidth]{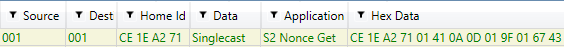}
\caption{Structure of manipulated S2 Nonce Get packet}
\label{fig:Nonce_S2}
\end{figure}

The only manipulation needed for the \emph{Nonce Get} frames is to change the source and destination addresses like in figure \ref{fig:Nonce_S0} or \ref{fig:Nonce_S2}. These are both changed to decimal 001. This is the address of the gateway itself. With the \emph{S2 Nonce Get} frames the sequence number is counted up as well. Now the gateway seems to send itself Nonce Get commands. Naturally the HomeID of the Z-Wave network has to be changed to the ID of the attacked one.


\begin{figure}[!h]
\includegraphics[width=\linewidth]{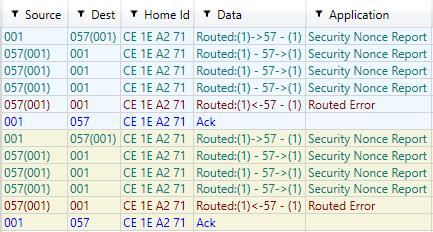}
\caption{Nonce Get S0 Packet reaction}
\label{fig:Nonce_S0_react}
\end{figure}

\begin{figure}[!h]
\includegraphics[width=\linewidth]{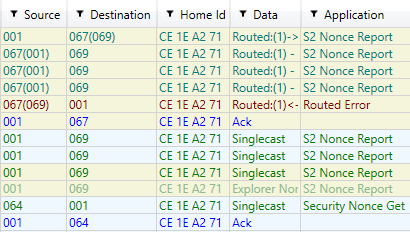}
\caption{Nonce Get S2 packet reaction}
\label{fig:Nonce_S2_react}
\end{figure}

As long as the attacker is sending these manipulated packets, the gateway tries to send Nonce Report packets to itself . This is not a problem in itself. Should there now be a node in the network available, which is capable of routing, it tries to route the Nonce Report packets via other routing nodes back to the gateway (see figure \ref{fig:Nonce_S0_react}), which doesn't recognize itself being the destination of the packet and tries the routing process several times over again.
This process takes a lot longer now, which in the meantime stops the gateway from processing received packets or commands sent via the smartphone app. As the gateway is the central managing entity within the network, which is responsible for both the logic and control of the connected nodes (i.e. the user's defined home automation programming), the whole Z-Wave network is blocked for as long as the gateway is blocked itself. Depending on the specific gateway and its implementation it might also be possible to use a not existing source address to get the gateway into the same state. During testing one such gateway was encountered among the test candidates (see figure \ref{fig:Nonce_S2_react}).

\subsection{Routed Nonce(nse)}
Through the previously mentioned effects a Denial of Service (DoS) attack is created if the attacker keeps on sending manipulated Nonce Get frames, as the network is constantly routing nonsense. Because this Denial of Service blocks the gateway, device internal automation aren't affected, like disassembly alarms. It is beneficial that the messages are sent unencrypted, which enables this attack for both security level S0 and S2. The efficiency of the attack varies though, depending on the specific manufacturer of the gateway. The most efficient result was a twenty minute DoS logjam against the targeted gateway utilizing only 256 sent packets.

\subsection{Automated Routed Noncense}
To automate this attack, a Python script was written to read the HomeID of the target network in reach, creating the manipulated messages automatically. Such created messages are then sent continuously to the targeted network. The intervals in which the packets are sent vary depending on the manufacturer of the gateway. If S2 Nonce Get packets are sent, a counter is used for the sequence number.

\subsection{Manipulating "Find nodes in range"}
\begin{figure}[h]
\includegraphics[width=\linewidth]{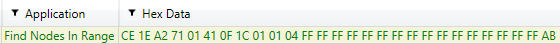}
\caption{Find Nodes In Range manipulated}
\label{fig:PowerofNOP_Package}
\end{figure}

The second DoS attack is realized through misuse of the \emph{Find Nodes In Range} command. Here, the addresses of the packet are changed again. Both source and destination addresses are set to decimal 001. Afterwards the payload is changed, filling it with 32 bytes of 0xFF (see figure \ref{fig:PowerofNOP_Package}). The payload usually depends on the known nodes in the network(assumed through testing). Because of this alteration, the gateway seems to send itself a \emph{Find Nodes In Range} packet. The HomeID of the attacked network has to be inserted as well. It works with both security level S0 and S2 without security level specific changes.

\subsection{Severity of the manipulation}

\begin{figure}[h]
\includegraphics[width=\linewidth]{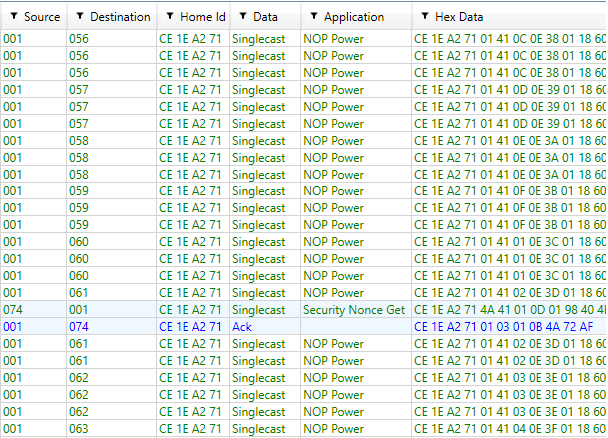}
\caption{Finde Nodes In Range Serverity}
\label{fig:PowerofNOP_Serv}
\end{figure}

Should this manipulated packet now been sent, the gateway gets the command to look for nodes in range. Therefore it sends \textit{NOP Power} packets to all possible addresses going several times through every address. It's waiting after each sent packet a fracture of time for an ACK and proceeds then with sending a packet to the next address. While the gateway executes this command, it doesn't answer or process any incoming messages as depicted in figure \ref{fig:PowerofNOP_Serv} the Nonce Get package or commands from the smartphone app. Therefore this can be used for an DoS attack as well. The execution of the manipulated packet takes the gateway a little under two minutes. The duration of the jamming is consistent through all manufacturers.
With these effects of a manipulated \emph{Find Nodes In Range} packet, a opportunity for a very efficient DoS attack is given. Since the gateway executes no other commands during the two minute time frame, the consecutively sent messages need to have perfect timing to guarantee a continuous Denial of Service on the gateway. Some old versions of the Z-Wave protocol can cause the gateway to send \emph{Command Complete} packets after the execution, which gives the attacker the exact moment to send another manipulated packet. This was the case with one of the tested gateways. The latest version of the Z-Wave protocol doesn't cause the gateway to send the \emph{Command Complete} packet. This version is obligatory for all manufacturers which require a S2 certification for their devices.

\subsection{Automated "Power of NOP(e)"}
The first part of the automation is the automatic insertion of the HomeID of the attacked network, which can be read-out of any normal packet sent within the Z-Wave network. The second part is the continuous sending of the Find Nodes In Range packets in given time intervals to jam the gateway and therefore the entire Z-Wave network. With the old implementations of the protocol there is the possibility to wait for an \emph{Command Complete} packet to determine the perfect moment for the attacker to send the next packet.

\section{Results}
There are two main differences between the tested smart home gateways. The first difference is the efficiency of the Routed Noncense attack. It probably has performance reasons how fast the gateway executes the routing measurements. The second difference is the used version of the Z-Wave protocol. 

There are still device for sale which implement older protocol releases from Silicon labs supporting only S0. These differ from the newer ones which support S2 as well. Therefore a gateway which had an older version only supporting S0 was easier to exploit by sending a fake \emph{Find Nodes In Range} packet, because it returned a \emph{Command Complete} packet after finishing the command. The newer versions of Z-Wave supporting S2 don't send this packet which makes it slightly harder to exploit. This can be used for perfect timing of the transmission of the next \emph{Find Nodes In Range} packets in an ongoing attack.
The jamming of a Z-Wave network can be useful for an intruder, who's goal is to break into a Z-Wave secured house without e.g. raising the alarm. In such attack scenario, the new found DoS attacks can be used instead of a normal hardware frequency jammer, which the difference that a jamming attack usually also effects neighbouring systems unintentionally. This attack can also be remote controlled if the attacker places a suitable SDR in a waterproof case powered by a battery pack and connected via GPRS or similar mobile protocols.
The DoS attacks can still be detected via heartbeat detection to an extend. If the heartbeat detection is executed by the gateway only, it maybe wouldn't be executed, similar to the commands coming from the smartphone application. Therefore it would have to detect that itself isn't sending heartbeats. On the other side a connected device would be able to detect a inactive gateway through heartbeats and would have to display the lost connection somehow. This would be difficult with window contacts or other devices which don't have a proper way to display these lost connections with other than a blinking LED, but not with alarm sirens or other security related devices which offer proper visual and audible feedback. As example the siren could activate a sound alarm. The user would have to be able to deactivate the heartbeat alarm in case there is a need to shut down the gateway or in the case of a downtime because of an update.

\section{Discussion}
In general this security issues were caused by mistakes of programmers during the implementation of the protocol. The \emph{Find Nodes In Range} command as example is not designed to be executed by the gateway. It is only supposed to be executed by other nodes during inclusion. The gateway is missing a rule to not send Nonce Report packets to non-existent addresses or itself as well. This shows how important prerequisites are, under which even less important or less used commands are allowed to be executed. 

\section{Conclusion}
Z-Wave is a radio protocol. Therefore it is always possible to execute a DoS attack with the use of a jammer. With this in mind the found DoS attacks are less severe, but it's a more concealed way of a DoS attack which isn't detectable though jamming detection using a Received Signal Strength Indicator. Besides that, it's a very efficient way to block the Z-Wave network. The main problem which was used for the attacks is the whole logic of the Z-Wave network being taken over by the gateway. Created automation, push messages and the processing of commands from the smartphone app are tasks of the gateway only. Because of that, the Z-Wave network has a star topology with some mesh capabilities when it comes to routing. So there is no need to block all nodes in the Z-Wave network if there is a way to block the central node. As an example the alarm siren wouldn't go off when the window contact senses the opening of the window, since the gateway wouldn't execute the automation for it. Both attacks take advantage of this by only blocking the gateway and therefore jamming the whole network. The attacks are both much more efficient than normal jamming and simple to pull off. The attack Power of NOPe requires the attacker to send one \emph{Find Nodes In Range} packet approx. every two minutes. The most efficient case of the Routed Noncense attack was able to block the gateway with just 256 packets for ca. 20 minutes. Both attacks can be repeated to block the gateway constantly over a intended time frame. Both attacks are only traceable if the attacked person has a Z-Wave sniffer active at all times. Then the person has to determine the unusual messages and would need to check if these were possible within Z-Wave. The source of the attack can't be determined, because the messages look like they've been sent from the gateway. A jammer would be much easier to be detected, because it blocks communication in itself. There are no permanent effects after the attacks and the gateway goes back to normal operation after the end of the attacks. Our written tool and the modified Scapy-Radio parts can be found in our GitHub repository: \url{https://github.com/A-Siemer/Dirtywave}

\section{Updates}
Silicon Labs acknowledged the vulnerabilities discovered in this paper and has already informed their customers via public announcement\cite{PSIRT-27}. Along with the announcement an updated Z-Wave implementation and protocol specification was made available, fixing the vulnerabilities which lead to the attack described above. The updated version of the Z-Wave protocol can be downloaded from the company's website\cite{Download_Z-Wave_SDK}.

\section{Acknowledgement}
We would like to thank Silicon Labs for their cooperation during the evaluation of the Z-Wave vulnerability found, especially
Jakob Buron. This project received funding from the Institute for Project Oriented Teaching (IPro-L) - University of Applied Sciences Emden/Leer.

\ifCLASSOPTIONcaptionsoff
  \newpage
\fi



%

\bibliographystyle{ieeetr}
\bibliography{paper} 

\begin{thebibliography}{10}

\bibitem{StatistaSHR2019}
Statista, ``{Smart Home Report 2019}.''
  \url{https://de.statista.com/statistik/studie/id/41155/dokument/smart-home-report/},
  2019.

\bibitem{BehrangGhanounBH2013}
S.~G. Behrang~Fouladi, ``Honey, i’m home!! - hacking z-wave home automation
  systems.'' Presentation at Black Hat-USA 2013, Las Vegas, NV, USA, July
  27–Aug. 1, 2013.

\bibitem{ZwaveSoC}
SiliconLabs, ``{EFR32ZG14 Z-Wave 700 Modem SoC DataSheet}.''
  \url{https://www.silabs.com/documents/public/data-sheets/efr32zg14-datasheet.pdf},
  01 2019.

\bibitem{ZWaveModule}
SiliconLabs, ``{ZGM130S Z-Wave 700 SiP Module DataSheet}.''
  \url{https://www.silabs.com/documents/public/data-sheets/zgm130s-datasheet.pdf},
  10 2019.

\bibitem{OpenZWave}
O.~Z-Wave, ``Open z-wave website.'' \url{http://www.openzwave.com/}.

\bibitem{ZWaveAllianceReport}
Z.-W. Alliance, ``2018 end of year z-wave ecosystem report.''
  \url{https://z-wavealliance.org/wp-content/uploads/2019/01/Z-Wave-Alliance-End-of-Year-Report-FINAL-for-web.pdf}.

\bibitem{HackRF}
{Great Scott Gadgets}, ``{HackRF One - SDR multi-frequency sender/receiver}.''
  \url{https://greatscottgadgets.com/hackrf/}.
\newblock Accessed: 2019-11-20.

\bibitem{GNU}
T.~G.~R. Foundation, ``{GNU Radio - tool for visual modelling software defined
  radio applications}.'' \url{https://www.gnuradio.org/}.
\newblock Accessed: 2019-11-20.

\bibitem{Scapy-Radio}
{Bastille Research}, ``{Scapy Radio Toolkit for Python}.''
  \url{https://github.com/BastilleResearch/scapy-radio}.
\newblock Accessed: 2019-11-20.

\bibitem{Pentestpartners}
PenTestPartners, ``{Z-Shave. Exploiting Z-Wave downgrade attacks}.''
  \url{https://www.pentestpartners.com/security-blog/z-shave-exploiting-z-wave-downgrade-attacks/}.
\newblock Accessed: 2019-11-20.

\bibitem{SecurityEvaluation}
B.~Fouladi and S.~Ghanoun, ``Security evaluation of the z-wave wireless
  protocol.''
  \url{https://pdfs.semanticscholar.org/10e1/21b903366ea81b94ca0c2e61c095cc087695.pdf},
  2013.
\newblock Blackhat Conference, Las Vegas, Nevada, USA.

\bibitem{BADENHOP2017112}
C.~W. Badenhop, S.~R. Graham, B.~W. Ramsey, B.~E. Mullins, and L.~O. Mailloux,
  ``The z-wave routing protocol and its security implications,'' {\em Computers
  \& Security}, vol.~68, pp.~112 -- 129, 2017.

\bibitem{spec}
SiliconLabs, ``{Z-Wave Specification}.''
  \url{https://www.silabs.com/products/wireless/mesh-networking/z-wave/specification}.
\newblock Accessed: 2019-11-20.

\bibitem{ZWaveApplication}
SiliconLabs, ``{SDS10865 - Z-Wave Application Security Layer (S0)}.''
  \url{https://www.silabs.com/documents/login/reference-manuals/SDS10865-Z-Wave-Application-Security-Layer-S0.pdf},
  03 2018.

\bibitem{Z-Wave-Transport-Encapsulation}
SiliconLabs, ``{SDS13783 - Z-Wave Transport-Encapsulation Command Class
  Specification}.''
  \url{https://www.silabs.com/documents/login/miscellaneous/SDS13783-Z-Wave-Transport-Encapsulation-Command-Class-Specification.pdf},
  10 2019.

\bibitem{findNodes}
SiliconLabs, ``{CTS10999 - Z-Wave Certification Test Specification}.''
  \url{https://www.silabs.com/documents/login/miscellaneous/CTS10999-Z-Wave-Certification-Test-Specification.pdf},
  03 2018.

\bibitem{Scapy}
P.~L. G.~P. Philippe~Biondi, Guillaume~Valadon, ``{Scapy - Python swiss army
  toolkit for packet manipulation}.'' \url{https://scapy.net/}.
\newblock Accessed: 2019-11-20.

\bibitem{Osmocom}
Osmocom, ``{OsmocomSDR}.''
  \url{https://osmocom.org/projects/gr-osmosdr/wiki/GrOsmoSDR}.
\newblock Accessed: 2019-11-20.

\bibitem{PSIRT-27}
SiliconLabs, ``{PSIRT-27 - Zwave Control Message DoS attack on Gateways and
  End-nodes}.'' internal customer security advisory/announcement, unpublished.

\bibitem{Download_Z-Wave_SDK}
SiliconLabs, ``{Z-Wave SDK}.''
  \url{https://www.silabs.com/products/development-tools/software/z-wave/embedded-sdk}.
\newblock Accessed: 2019-11-20.

\end{thebibliography}

%








\end{document}